\documentclass[12pt]{article}
\pdfoutput=1
\usepackage{epsf,amsfonts,amssymb,epsfig,amsmath,mathtools,graphics,slashed}
\usepackage{hyperref,hep,graphicx,subfigure}
\makeatletter

\newcommand{\Rmnum}[1]{\expandafter\@slowromancap\romannumeral #1@}
\makeatother

\newcommand{\nn}{\notag \\}

\begin{document}

\makeatletter
\renewcommand{\theequation}{\thesection.\arabic{equation}}
\@addtoreset{equation}{section}
\makeatother

\baselineskip 18pt

\begin{titlepage}

\vfill

\begin{flushright}
Imperial/TP/2013/JG/02\\
\end{flushright}

\vfill

\begin{center}
   \baselineskip=16pt
   {\Large\bf On the thermodynamics of\\
   periodic AdS black branes}
  \vskip 1.5cm
     Aristomenis Donos and Jerome P. Gauntlett\\
      \vskip .6cm
      \begin{small}
      \textit{Blackett Laboratory, 
        Imperial College\\ London, SW7 2AZ, U.K.}
        \end{small}\\*[.6cm]

\end{center}

\vfill

\begin{center}
\textbf{Abstract}
\end{center}

\begin{quote}
We consider asymptotically AdS black brane solutions that are dual to
CFTs with periodic dependence on the spatial directions, arising 
from either a spontaneous or an explicit breaking of translational symmetry.
We derive a simple expression for the variation of the free-energy with respect to
changing the periods. This explains some observations, based on numerics, that have arisen 
in the explicit construction of thermodynamically preferred black holes in the case that the 
spatial directions are infinite in extent and the symmetry is spontaneously broken.
It also leads to Smarr-type relations involving the boundary stress tensor.
\end{quote}

\vfill

\end{titlepage}
\setcounter{equation}{0}


\section{Introduction}
Strongly coupled field theories with a dual holographic description can exist in a wide variety of phases 
described by novel black hole solutions. A particularly interesting class of examples are CFTs
in flat spacetime which spontaneously break translation invariance when 
held at finite chemical potential with respect to a global $U(1)$ symmetry. For example, asymptotically $AdS_5$ black holes 
corresponding to $d=4$ CFTs acquiring helical current phases and helical superconducting phases have been
studied in \cite{Nakamura:2009tf,Donos:2012wi} and \cite{Donos:2011ff,Donos:2012gg}, respectively 
(see also \cite{Domokos:2007kt,Iizuka:2012iv}). Similarly, asymptotically
$AdS_4$ black holes corresponding to $d=3$ CFTs acquiring striped phases have been studied in 
\cite{Donos:2011bh,Rozali:2012es,Donos:2013wia,Withers:2013loa,Withers:2013kva,Rozali:2013ama} (see also \cite{Bergman:2011rf}). 

In these particular examples, the spatial modulation is confined to a single spatial direction and is periodic. The corresponding 
black hole solutions depend
on both the temperature $T$ and a wave-number $k$, which specifies the period $L$ of the spatial modulation via $L=2\pi/k$.
When the relevant spatial direction is non-compact one is interested in minimising the free-energy density with
respect to $k$ in order to obtain the thermodynamically preferred black holes. In the detailed numerical 
constructions carried out in \cite{Donos:2012gg,Donos:2012wi,Withers:2013kva,Rozali:2013ama} 
this variation turned out to impose simple conditions on the boundary data, but
the underlying reason for this was obscure.

Here we will provide an explanation for these results and moreover, in doing so, obtain some more general results with wider applicability.
The arguments involve simple extensions of basic holographic results including that the free-energy is given by the 
on-shell action and that stress tensor is obtained as an on-shell variation of the action with respect to the boundary metric.
Nevertheless, our results are of significant practical utility. Specifically, to obtain the thermodynamically preferred black holes in
the examples mentioned above, one can now simply construct the one parameter family of black hole solutions, labelled by the temperature $T$, with the preferred value of
$k$ given, implicitly, as a function of $T$, rather than construct a two-parameter family of black holes
specified by $T,k$ and then minimise the free energy with respect to $k$, as was done hitherto.

As mentioned, our results also have more general applicability. Firstly, the periodic spatial modulation can be in all spatial directions.
Secondly, it covers cases where the translation symmetry is
explicitly broken by source terms, such as periodic boundary chemical potentials, currents or metrics.
Such examples have been studied in a variety of situations in, for example, \cite{Aperis:2010cd,Flauger:2010tv,Maeda:2011pk,Horowitz:2012ky,Horowitz:2012gs,Donos:2012js,Horowitz:2013jaa}.
Our results are also applicable to the case when the spatial directions are compact with specific period. 
In particular, we obtain a compact expression for the variation of the free-energy with respect to changing the period of the compact direction.
Finally, we will obtain simple Smarr-type formulae and constraints involving the boundary stress tensor. 
A different approach to Smarr-type formulae in the context of planar AdS
black holes has recently been studied in \cite{El-Menoufi:2013pza,El-Menoufi:2013tca} and we will see how 
our results generalise those obtained there.

We will only analyse in detail the case of gravity coupled to a gauge-field, since it is straightforward to 
include other matter fields. We prove our main results in section \ref{result}.
We discuss the example of $D=5$ Einstein-Maxwell-Chern Simons theory in some detail in section \ref{cs} since it provides a satisfying realisation of the arguments. In section \ref{othex} we briefly discuss some other examples, which are somewhat simpler. We conclude in section \ref{fincom} by briefly discussing generalisations including the
addition of external magnetic fields.

\section{Main Results}\label{result}
Consider a theory in $D=d+1$ bulk spacetime dimensions which couples gravity to an abelian gauge-field. 
We are interested in stationary black hole solutions that asymptotically approach $AdS_D$
with $d$-dimensional boundary metric given by
\begin{align}\label{bone}
ds_{B}^{2}&=\gamma_{\mu\nu}\,dx^{\mu}\,dx^{\nu}\,,
\end{align}
with $\mu,\nu=1, \dots, d$. It will be helpful to split the boundary coordinates into time and space coordinates $x^\mu=(t,x^i)$ 
with $i=1,\dots, d-1$. 
The black hole is taken to have a Killing horizon generated by a bulk Killing vector which approaches $\partial_t$ on the boundary.
We assume that $\gamma_{\mu\nu}$ are periodic functions of the globally defined $x^i$ only, with period $L_i$. This covers both the ``compact case" when 
the spatial coordinates $x^i$ are of finite extent, with $x^i\cong x^i+L_i$, and the ``non-compact case" of periodic configurations when the $x^i$ have 
infinite extent.
In the Euclidean continuation of the black hole space-time we should take
$t\to -i\tau$ and the co-ordinate $\tau$ has period $\Delta \tau=T^{-1}$ where $T$ is the temperature of the dual CFT.
The gauge-field is taken to asymptotically approach the boundary behaviour given by
\begin{align}\label{btwo}
A_B&=a_{\mu}\,dx^{\mu}\,,
\end{align}
with $a_\mu$ periodic functions of $x^{i}$ only. 

It is worth noting that this set-up includes cases where the boundary 
metric is taken to be flat, $\gamma_{\mu\nu}=\eta_{\mu\nu}$, with $a_t\equiv \mu$, where $\mu$ is a constant, and $a_{x^i}=0$. This 
corresponds to studying CFTs in flat spacetime when held at finite temperature and chemical potential $\mu$.
In particular, the periodic dependence of the bulk fields on the coordinates $x^i$ arises because there is a 
spontaneous breaking of translation invariance.
More generally, though, the asymptotic boundary conditions in \eqref{bone} and \eqref{btwo} 
also allow the ``sources" $\left\{\gamma_{\mu\nu},a_{\mu} \right\}$ to have non-trivial dependence on the 
periodic coordinates corresponding to an explicit breaking of translation invariance.

In the case that the $x^i$ are of finite extent we are interested in the free-energy, which is a functional defined by
\begin{align}
W&=W\left(\gamma_{\mu\nu},a_{\mu};\Delta\tau,L_{i}\right)=\frac{1}{\Delta\tau}\,I_{OS}\left(\gamma_{\mu\nu},a_{\mu};\Delta\tau,L_{i}\right)\,,
\end{align}
where $I_{OS}$ is the on-shell value of the {\it total} Euclidean action, including boundary terms,
integrated over a period of $\tau$ and the $x^{i}$. 
When the $x^i$ have infinite extent we are interested in the free-energy density
defined by 
\begin{align}
w&=w\left(\gamma_{\mu\nu},a_{\mu};\Delta\tau,L_{i}\right)=\frac{1}{\Delta\tau\,\Pi}\,I_{OS}\left(\gamma_{\mu\nu},a_{\mu};\Delta\tau,L_{i}\right)\,,
\end{align}
where $\Pi=\prod_{i=1}^{d-1}L_{i}$. 

To begin our derivation we consider an on-shell variation of the total action $I$
with respect to the asymptotic data $\left\{\gamma_{\mu\nu},a_{\mu} \right\}$ holding fixed the
periods $\Delta \tau$ and $L_i$. 
We assume that the variation $\left\{\delta\gamma_{\mu\nu},\delta a_{\mu} \right\}$ has the same periodicity as
$\left\{\gamma_{\mu\nu},a_{\mu} \right\}$.
Using standard AdS/CFT results \cite{Balasubramanian:1999re} we have:
\begin{align}\label{firstvar}
\tilde \delta I_{OS}&=-\,\int_{0}^{\Delta\tau} \int_{0}^{\{L_{i}\}}\,d\tau d^{d-1}x\,\sqrt{-\det{\gamma}}\,\left(\tfrac{1}{2}T^{\mu\nu}\,\delta \gamma_{\mu\nu}+J^{\mu}\,\delta a_{\mu} \right)\,,
\end{align}
where $T^{\mu\nu}$ is the stress-energy tensor and $J^\mu$ is the current density of the boundary CFT. 
Note that the integrand is independent of $\tau$ and hence the integral over $\tau$ just gives a factor of $\Delta\tau$.

To obtain the variation with respect to the periods we proceed as follows. 
By a simple change of variables
\begin{align}\label{eq:scale_coords}
\tau=\Delta\tau\,\tilde{\tau},\qquad x^{i}=L_{i}\,\tilde{x}^{i}\,,
\end{align}
we have that
\begin{align}
I_{OS}\left(\gamma_{\mu\nu},a_{\mu};\Delta\tau,L_{i}\right)=I_{OS}\left(\tilde{\gamma}_{\mu\nu},\tilde{a}_{\mu};1,1\right)\,,
\end{align}
with
\begin{align}
ds_{B}^{2}&=\tilde{\gamma}_{\mu\nu}\,d\tilde{x}^{\mu}\,d\tilde{x}^{\nu}\,,\nn
&\equiv\gamma_{\tau\tau}\left(\Delta\tau \right)^{2}\,d\tilde{\tau}^{2}+2\gamma_{\tau x^i}\Delta\tau L_i d\tilde \tau d\tilde x^i+\gamma_{x^ix^j}L_iL_j d\tilde x^i d\tilde x^j\,,
\nn
A_B&=a_\tau\,\Delta\tau\,d\tilde{\tau}+a_{x^i}\,L_{i}\,d\tilde{x}^{i}\,.
\end{align}
Notice that the functions $\left\{\gamma_{\mu\nu},a_{\mu} \right\}$ have period equal to one with respect to the rescaled coordinates 
$(\tilde \tau,\tilde x^i)$. Using $\sqrt{\det{\tilde\gamma}}=\Delta\tau\Pi\sqrt{\det{\gamma}}$
and the chain rule, the total variation is now seen to be
\begin{align}\label{eq:total_variation}
\delta I_{OS}&=\quad\tilde \delta I_{OS}\notag\\
&-\Delta\tau\,\Pi\,\int_{0}^{1}d^d\tilde{x}\sqrt{\det{\gamma}}\,\left(\tfrac{1}{2}\,T^{\tau \tau }\gamma_{\tau \tau }\,\frac{\delta\left((\Delta\tau)^{2} \right)}{(\Delta\tau)^{2}}+T^{\tau x^i}\gamma_{\tau x^i}\,\frac{\delta\left(\Delta\tau\,L_{i} \right)}{\Delta\tau\,L_{i}} +\tfrac{1}{2}\,T^{x^ix^j}\gamma_{x^ix^j}\,\frac{\delta\left( L_{i}\,L_{j}\right)}{L_{i}L_{j}}\right)\notag\\
&-\Delta\tau\,\Pi\,\int_{0}^{1}d^d\tilde{x}\sqrt{\det{\gamma}}\,\left(J^{\tau }a_{\tau }\, \frac{\delta\left(\Delta\tau \right)}{\Delta\tau}+J^{x^i}a_{x^i}\,\frac{\delta L_{i}}{L_{i}}\right)\,,
\end{align}
where we also used e.g. $\tilde T^{\tau \tau }=T^{\tau \tau }/(\Delta \tau)^2$.
After reorganising \eqref{eq:total_variation} and recalling that $\delta\Delta\tau/ \Delta\tau=-\delta T/T$ we obtain the result
\begin{align}
&\delta I_{OS}=\quad\tilde \delta I_{OS}
+\frac{\delta T}{T}\,\int_{0}^{\Delta\tau}\int_{0}^{\{L_{i}\}}d\tau d^{d-1}x\sqrt{-\det{\gamma}}\,\left(T^{\tau\tau}\gamma_{\tau\tau}+T^{\tau x^i}\gamma_{\tau x^i}+J^{\tau }a_\tau  \right)\notag\\
&-\sum_{i,j}\frac{\delta L_{i}}{L_{i}}\,\int_{0}^{\Delta\tau}\int_{0}^{\{L_{i}\}}d\tau d^{d-1}x\sqrt{-\det{\gamma}}\,\left(T^{\tau x^i}\gamma_{\tau x^i}+T^{x^ix^j}\gamma_{x^ix^j}+J^{x^i}a_{x^i} \right)\,,
\end{align}
where the integrals
over $\tau$ just give a factor of $\Delta\tau$.

Thus, the variation of the free-energy is given by
\begin{align}\label{eq:final_var}
\delta W&=- \int_{0}^{\{L_{i}\}}d^{d-1}x\,\sqrt{-\det{\gamma}}\,\left(\tfrac{1}{2}T^{\mu\nu}\,\delta \gamma_{\mu\nu}+J^{\mu}\,\delta a_{\mu} \right)
-S{\delta T}\notag\\
&-\sum_{i,j}\frac{\delta L_{i}}{L_{i}}\left[\int_{0}^{\{L_{i}\}}d^{d-1}x\sqrt{-\det{\gamma}}\,\left(T^{tx^i}\gamma_{tx^i}+T^{x^ix^j}\gamma_{x^ix^j}+J^{x^i}a_{x^i} \right)\right]\,,
\end{align}
where the entropy $S$ is defined as
\begin{align}\label{strop}
S\equiv-\frac{1}{T}\left[W+\,\int_{0}^{\{L_{i}\}}d^{d-1}x\sqrt{-\det{\gamma}}\,\left(T^{tt}\gamma_{tt}+T^{tx^i}\gamma_{tx^i}+J^{t}a_{t} \right)\right]\,,
\end{align}
and we have switched back to the Lorentzian components in the integrand.
The entropy $S$ can be related to geometric properties of the black hole by making some assumptions about the bulk
gravitational theory as, for example, in \cite{Papadimitriou:2005ii}. In particular, without higher derivative terms, it is
given by one quarter of the area of the event horizon divided by Newton's constant, as usual.

Similarly the variation of the free-energy density is given by
\begin{align}\label{eq:final_var2}
\delta w&=
-\Pi^{-1}\int_{0}^{\{L_{i}\}}d^{d-1}x\,\sqrt{-\det{\gamma}}\,\left(\tfrac{1}{2}T^{\mu\nu}\,\delta \gamma_{\mu\nu}+J^{\mu}\,\delta a_{\mu} \right)-s{\delta T}\notag\\
&-\sum_{i,j}\frac{\delta L_{i}}{L_{i}}\left[w+\Pi^{-1}\,\int_{0}^{\{L_{i}\}}d^{d-1}x\sqrt{-\det{\gamma}}\,\left(T^{tx^i}\gamma_{tx^i}+T^{x^ix^j}\gamma_{x^ix^j}+J^{x^i}a_{x^i} \right)\right]\,,
\end{align}
where $s\equiv S/\Pi$ is the entropy density. 
In the non-compact case, and when there is no explicit breaking of the translation symmetry in
the $i$th direction,
the thermodynamically preferred periodic configurations will minimise the free-energy density with respect to
the wave numbers $k_i\equiv 2\pi/L_i$ for each $i$. From \eqref{eq:final_var2} we deduce that
for each such $i$
\begin{align}
\label{freeda}
w&=-\Pi^{-1}\,\int_{0}^{\{L_{i}\}}d^{d-1}x\sqrt{-\det{\gamma}}\,\left(T^{tx^i}\gamma_{tx^i}+\sum_jT^{x^ix^j}\gamma_{x^ix^j}+J^{x^i}a_{x^i} \right)\,,\quad \text{no sum on $i$}\,.
\end{align}
and we emphasise that this applies even if there is no dependence on the spatial coordinates (and hence is
also valid in the compact case). As we will illustrate later, when \eqref{freeda} is combined
with the expression for the free-energy density arising from \eqref{strop},
\begin{align}\label{wst}
w&=-Ts-\Pi^{-1}\int_{0}^{\{L_{i}\}}d^{d-1}x\sqrt{-\det{\gamma}}\,\left(T^{tt}\gamma_{tt}+T^{tx^i}\gamma_{tx^i}+J^{t}a_{t} \right)\,,
\end{align}
we can obtain at least one Smarr-type formulae.
These results generalise\footnote{In obtaining \eqref{freeda} we implicitly 
assumed that the Killing vector $\partial_i$ acts with no fixed points 
in the bulk space-time. Indeed when it does, the period of the coordinate $x^i$ is fixed by the regularity of
the bulk spacetime at the fixed point set and hence it cannot be freely varied. In the situation when $\partial_\tau$ acts freely
and $\partial_{x^i}$ does not, as for example in the AdS soliton solution of \cite{Witten:1998zw},
we can obtain anologous results with suitable modifications. For example, \eqref{freeda} will 
have an extra $-T_Bs_B$ term on the right hand side, where $T_B=1/L_i$ is the ``temperature" of the fixed point ``bubble" and, without higher derivative gravity, $s_B$ is 1/4 of the area of the bubble divided
by $\Pi L_i^{-1}\Delta \tau$ times Newton's constant. Such terms were considered in \cite{El-Menoufi:2013pza} building on \cite{Kastor:2008wd}.
 } 
those of \cite{El-Menoufi:2013pza}.

The Euclidean boundary is a $d$-dimensional torus parametrised by the globally defined periodic coordinates $\tau,x^i$. The circle with
tangent vector $\partial_\tau$ is singled out by the presence of the black hole in the bulk spacetime. 
It is illuminating to consider alternative global coordinates, related by $Sl(d,\mathbb{Z})$ transformations, that preserve the Killing vector $\partial_\tau$. 
For example, consider
\begin{align}
\tau&=\bar \tau+\alpha_i\bar x^i,\quad \text{no sum on $i$}\nn
x^i&=\bar x^i
\end{align}
with $\partial_{\bar \tau}=\partial_\tau$ and $\partial_{\bar x^i}=\partial_{x^i}+\alpha_i\partial_\tau$. We choose $\alpha_i=\Delta\tau/\Delta x^i$
so that the torus is parametrised by periodic coordinates $(\bar\tau, \bar x^i)$ with the same periods as $(\tau,x^i)$.
Now suppose we had carried out calculation leading to \eqref{eq:final_var2} using the barred coordinates. By rewriting the various tensors in
the unbarred coordinates we would find the expression \eqref{eq:final_var2} plus another term proportional to $\alpha_i$. Consistency implies that
this latter term should vanish and hence we conclude that for each $i$ we have
\begin{align}\label{consone}
\int_{0}^{\{L_{i}\}}d^{d-1}x\sqrt{-\det{\gamma}}\left(T^{x^it}\gamma_{tt}+\sum_k T^{x^ix^k}\gamma_{x^k t}+J^{x^i} a_t\right)=0\,.
\end{align}
We can also consider similar $Sl(d-1,\mathbb{Z})$ transformations only involving the spatial coordinates, which
clearly leave $\partial_\tau$ unchanged. A similar calculation implies that
\begin{align}\label{constwo}
\int_{0}^{\{L_{i}\}}d^{d-1}x\sqrt{-\det{\gamma}}\left(T^{x^it}\gamma_{tx^j}+\sum_k T^{x^ix^k}\gamma_{x^k x^j}+J^{x^i} a_{x^j}\right)=0,\qquad i\ne j\,.
\end{align}

\section{Black holes dual to helical current phases}\label{cs}
We review the construction of $D=5$ black holes dual to helical current phases that
were constructed in \cite{Donos:2012wi}, building on \cite{Nakamura:2009tf}, 
and show how they satisfy the results of the previous
section. In particular, we will obtain an understanding of how the
thermodynamically preferred black holes in the non-compact case found numerically in 
\cite{Donos:2012wi} are simply specified
by boundary data.

\subsection{Summary of some results of \cite{Donos:2012wi}}
The $D=5$ action is given by
\begin{align}
S=S_{bulk}+S_{bdy}\,,
\end{align}
with the bulk action given by
\begin{align}\label{eq:lagemcs}
S_{bulk}&=\int d^5 x\sqrt{-g}\left[(R+12)-\frac{1}{4} F_{\mu\nu}F^{\mu\nu}\right]-\frac{\gamma}{6}\int F\wedge F \wedge{A}\,,
\end{align}
where $\gamma$ is a constant, and the boundary action $S_{bdy}$ is \cite{D'Hoker:2009bc}.
\begin{align}
S_{bdy}&=\lim_{r\to\infty}\int d^4x\sqrt{-g_\infty}\left(2K-6+\frac{1}{2}R_\infty+\frac{1}{4}\ln r F_{mn}F^{mn}\right)\,,
\end{align}
where $g_\infty$ is the induced boundary metric with Ricci scalar $R_\infty$ and $K$ is the extrinsic curvature. 
The ansatz for the helical black holes is given by 
\begin{align}\label{eq:ansatz}
ds^{2}&=-g\,f^{2}\,dt^{2}+\frac{dr^{2}}{g}+h^{2}\,\omega_{1}^{2}+r^{2}e^{2\alpha}\left(\omega_{2}+Qdt\right)^2+r^2e^{-2\alpha}\,\omega_{3}^{2}\,,\notag\\
A&=a\,dt+b\,\omega_2\,,
\end{align}
where $f$, $g$, $h$, $\alpha$, $Q$, $a$ and $b$ are functions of the radial coordinate $r$ only.
Furthermore, the $\omega_i$ are the left-invariant Bianchi $VII_0$ one-forms given by
\begin{align}\label{eq:one_forms}
&\omega_{1}=dx^{1},\nn
&\omega_{2}=\cos\left(kx^{1}\right)\,dx^{2}-\sin\left(kx^{1}\right)\,dx^{3},\nn
&\omega_{3}=\sin\left(kx^{1}\right)\,dx^{2}+\cos\left(kx^{1}\right)\,dx^{3}\,.
\end{align}
The axis of the helical symmetry is along the $x^1$ direction and has period $L_1=2\pi/k$.

As $r\to\infty$ the black holes asymptotically approach $AdS_5$:
\begin{align}\label{mubexpmet}
g&=r^2 (1-\frac{M}{r^4}+\dots),\qquad\qquad
f=1+\frac{-c_h+\frac{k^2 \mu_b^2}{48}}{r^4}
+\dots\,,\nn
 h&=r \Big(1+\frac{c_h}{r^4}+\dots),\qquad
   \alpha=\frac{c_\alpha}{r^4}+\dots\,,\qquad\qquad
 Q=\frac{c_Q}{r^4}+\dots\,,\nn
  a&=\mu+\frac{q}{r^2}+\dots,\qquad\qquad\qquad
  b=\mu_b+\frac{c_b}{r^2}+\dots\,.
 \end{align}
Observe that the asymptotic boundary metric is flat, $\gamma_{\mu\nu}=\eta_{\mu\nu}$. We also observe that the asymptotic gauge-field
is $A_B=\mu dt+\mu_b \omega_2$, where $\mu$ is a constant chemical potential and $\mu_b$ parametrises
an explicit helical current source. {\it For the case of spontaneous generation of helical current phases one should set $\mu_b=0$}.
Indeed the explicit black hole solutions that were constructed numerically in \cite{Donos:2012wi} had $\mu_b=0$. The case of
$\mu_b\ne 0$ was also considered in \cite{Donos:2012js}.
At the black hole horizon, located at $r=r_{+}$,  the functions have the analytic expansion
\begin{align}\label{nhexp}
g&=g_{+}\,\left(r-r_{+}\right)+\dots,\qquad
f=f_{+}+\dots\,,\notag\\
h&=h_{+}+\dots,\qquad
\alpha=\alpha_{+}+\dots,\quad
Q=Q_{+}(r-r_+)+\dots\,,\notag\\
a&=a_{+}\,\left(r-r_{+}\right)+\dots,\qquad
b=b_{+}+\dots\,.
\end{align}

The free-energy for such black hole solutions was calculated in \cite{Donos:2012wi} by evaluating the total 
on-shell euclidean action. Two expressions for the on-shell {\it bulk} action were obtained:
\begin{align}
[I_{bulk}]_{OS}&=vol_3\Delta\tau\int_{r_+}^\infty dr \xi_1'\,,\nn
&=vol_3\Delta\tau\int_{r_+}^\infty \xi_2'\,,
\end{align}
where
\begin{align}\label{xiexps}
\xi_1&=2rghf+\frac{r^{4}e^{2\alpha}h}{2f}QQ^{\prime}+\frac{1}{2}he^{-2\alpha}fgbb^{\prime}+\frac{1}{2f} r^{2}h\left(a^{\prime}-Qb^{\prime} \right)bQ +\frac{1}{6}k\gamma ab^{2}\,,\nn
\xi_2&=r^{2}hfg^{\prime}+2r^{2}hgf^{\prime} -\frac{h}{f}r^{4}e^{2\alpha}QQ^{\prime}-\frac{1}{f}r^{2}ha \left(a^{\prime}-Qb^{\prime} \right)-\frac{1}{3}k\gamma ab^{2}\,,
\end{align}
leading to 
\begin{align}\label{osact}
w&=-M-\mu_b\,c_b-\frac{1}{12}\mu_b^{2}k^{2}+\frac{1}{6}\mu\mu_b^{2}k\gamma\,,\nn
 &=3M+8c_{h} +2\mu q-Ts-\frac{1}{8}\mu_b^{2}k^{2}-\frac{1}{3}\mu \mu_b^{2}k\gamma\,,
 \end{align}
with the equality implying a Smarr-type formula.

Similarly the boundary stress tensor was evaluated with the result:
\begin{align}\label{emtvev2}
 T_{tt}&=3M+8c_{h}-\frac{1}{8}\,\mu_b^{2}k^{2},\qquad
 T_{tx^{2}}=\,4c_{Q}\,\cos\left(k x^{1}\right),\qquad
 T_{tx^{3}}=-4c_{Q}\,\sin\left(k x^{1}\right),\notag\\
 T_{x^{1}x^{1}}&=M+8c_{h}-\frac{7}{24}\,\mu_b^{2}k^{2},\qquad
 T_{x^{2}x^{3}}=-(8c_{\alpha}+\frac{1}{8}\,\mu_b^{2}k^{2})\,\sin\left(2kx^{1}\right)\,,\nn
 T_{x^{2}x^{2}}&=M -\mu_b^2 k^2/6+(8c_{\alpha}+\frac{1}{8}\,\mu_b^{2}k^{2})\,\cos\left(2kx^{1} \right)\,, \notag\\
 T_{x^{3}x^{3}}&=M-\mu_b^2 k^2/6 -(8c_{\alpha}+\frac{1}{8}\,\mu_b^{2}k^{2})\,\cos\left(2kx^{1} \right),\quad
\end{align}
where we have corrected some typos\footnote{Note also that the $T^\mu{}_{\mu}=-\mu_b^2k^2/2=-\tfrac{1}{4}F_{\mu\nu}F^{\mu\nu}$, correcting another typo in \cite{Donos:2012wi}.} 
 in the expressions for $T_{x^{2}x^{2}}$ and $T_{x^{3}x^{3}}$ that were given in \cite{Donos:2012wi}.
The boundary current is\footnote{Note that there is a sign difference in the definition of the current density compared to \cite{Donos:2012wi}.}:
\begin{align}\label{vevofJ2}
J_{t}&=2q-\frac{1}{3}\mu_b^{2}k\gamma,\qquad\qquad
 J_{x^1}=0\,,\notag\\
 J_{x^2}&=(2c_b+\frac{1}{2}\mu_bk^{2}-\frac{1}{3}\mu \mu_bk\gamma )\cos(kx^1)\,,\nn
 J_{x^3}&=-(2c_b+\frac{1}{2}\mu_bk^{2}-\frac{1}{3}\mu \mu_bk\gamma )\sin(kx^1)\,.
\end{align}
The variation of the free-energy, {\it holding $k$ fixed}, was shown to be given by
\begin{align}\label{varwgen}
\delta w=\left(2q-\frac{1}{3}\mu_b^{2}k\gamma \right)
\,\delta\mu-\left(2c_b+\frac{1}{2}\mu_bk^{2}-\frac{1}{3}\mu\mu_bk\gamma \right)\delta\mu_b-s \delta T\,.
\end{align}
In addition the derivative of the total Euclidean action with respect to $k$ was given in \cite{Donos:2012wi} implying that
\begin{align}\label{xvarhp}
k\,\frac{\delta w}{\delta k}=\int_{r_+}^\infty dr\bigg(&\frac{e^{2\alpha}k^{2}fb^{2}}{h}+\frac{4k^{2}r^{2}f}{h}\,\sinh^{2}(2\alpha)-\frac{e^{-2\alpha}k^{2}r^{4}Q^{2}}{hfg}\nn&
-\frac{1}{3}k\gamma b^{2}a^{\prime}+\frac{1}{3}k\gamma b b^{\prime}a\Bigg)_{OS}-\lim_{r\to\infty}\left(\ln r\frac{fg^{1/2}e^{2\alpha}b^2}{h}k^2\right)_{OS}\,,
\end{align}
where the $OS$ refers to the right-hand side being evaluated on-shell and the last term arises from a contribution from
the boundary terms.
Finally, for the explicit black hole solutions that were constructed in \cite{Donos:2012wi}, which was for the case when $\mu_b=0$ and the translation symmetry was broken spontaneously,
it was shown from the numerics that the vanishing of \eqref{xvarhp} implied simply that $c_h=0$; the reason for this was unclear.

\subsection{Connection with new results}
Given that the boundary metric $\gamma_{\mu\nu}=\eta_{\mu\nu}$, the conditions \eqref{eq:final_var2}-\eqref{wst}
read:
\begin{align}\label{twomainex}
\delta w
&=-\frac{k}{2\pi}\int_0^{2\pi/k} dx^1 J^\mu\delta a_\mu-s \delta T+\frac{\delta k}{k}\left(w+\frac{k}{2\pi}\int_0^{2\pi/k} dx^1 T^{x^1x^1}\right)
\\
\label{sndone}
w&=-\frac{k}{2\pi}\int_0^{2\pi/k} dx^1 (T^{x^ix^i}+J^{x^i} a_{x^i}),\qquad\text{no sum on $i$, $\qquad$$i=2,3$}\,,\\
w&=-Ts+\frac{k}{2\pi}\int_{0}^{2\pi/k}dx^1\left(T^{tt}-J^{t}a_{t} \right)\,,
\label{wsteg}
\end{align}
where we have used $L_1=2\pi/k$ and hence $\delta L_1/L_1=-\delta k/k$. The boundary 
gauge-field data is given by
\begin{align}\label{delper}
a_\mu&= (\mu,0,\mu_b\cos(kx^1),-\mu_b\sin(kx^1))\,,\nn
\delta a_\mu&=(\delta\mu,0,\delta\mu_b\cos(kx^1),-\delta\mu_b\sin(kx^1))\,.
\end{align}
Note that the expression for the 
free-energy given in \eqref{sndone} arises from the fact that there is no spatial modulation in the $x^2$ and $x^3$ directions and is therefore valid when the $x^2,x^3$ directions are either compact or non-compact. In the case that
the $x^1$ direction is non-compact, and when there is no explicit breaking of the translation
symmetry i.e. $\mu_b=0$, then we should also impose $\delta w/\delta k=0$ and this leads to another expression for the free-energy.

We now substitute the expressions for the boundary stress tensor \eqref{emtvev2} and current density \eqref{vevofJ2}
that were given in the last subsection and show how \eqref{twomainex}-\eqref{wsteg} imply various other conditions that
were summarised in the last subsection.

Starting with \eqref{wsteg}, after substituting the expressions for $T^{tt}$ and $J^t$ from \eqref{emtvev2}, \eqref{vevofJ2},
we immediately obtain the expression for $w$ given in the second line of \eqref{osact}. We
next consider \eqref{sndone}. After substituting  the expressions for $T^{x^ix^i}$ and $J^{x^i}$ from \eqref{emtvev2}, \eqref{vevofJ2}
and carrying out the integral over $x^1$, for both $i=2$ and $i=3$ we obtain the expression for $w$
given in the first line of \eqref{osact}. In particular, \eqref{sndone} and \eqref{wsteg} provide a useful technique to derive Smarr-type formula just using
the expressions for the boundary stress tensor and current.

We now consider \eqref{twomainex}. We first focus on holding $k$ fixed. Using the expression
for $J^\mu$ given in \eqref{vevofJ2} we have\footnote{If one considers perturbations of the form
$\delta a_\mu=(0,0,\delta\mu^{(n)}_b\cos(nkx^1),-\delta\mu^{(n)}_b\sin(nkx^1))$ for some integer $n\ne \pm1$, 
which also respect the periodicity of the background, one finds that they give no contribution to $\delta w$ in \eqref{twomainex} after integrating over $x^1$.}
\begin{align}
J^\mu\delta a_\mu=-\left(2q-\frac{1}{3}\mu_b^{2}k\gamma \right)
\,\delta\mu
+\left(2c_b+\frac{1}{2}\mu_bk^{2}-\frac{1}{3}\mu\mu_bk\gamma \right)\delta\mu_b\,,
\end{align}
which is independent of $x^1$, and we recover \eqref{varwgen}.

Next, we consider the terms in \eqref{twomainex} involving $\delta k$.
Substituting in the expression for $T^{x^1x^1}$ given in \eqref{emtvev2} we deduce that
\begin{align}\label{ldg}
k\frac{\delta w}{\delta k} 
&=w+M+8c_{h}-\frac{7}{24}\,\mu_b^{2}k^{2}\,,\nn
&=8c_h-\mu_b\,c_b-\frac{3}{8}\mu_b^{2}k^{2}+\frac{1}{6}\mu\mu_b^{2}k\gamma\,,
\end{align}
where the second line follows from the first line of \eqref{osact}, which in turn we have just shown is equivalent to \eqref{sndone}. 
Thus, when $x^1$ is a non-compact direction and we have no explicit breaking of translation symmetry i.e. $\mu_b=0$, the thermodynamically preferred black holes must obey $\frac{\delta w}{\delta k}=0$ and hence satisfy 
$c_h=0$, explaining the numerical results observed in \cite{Donos:2012wi}.

The result \eqref{twomainex} also implies that the integrand on the right hand side of \eqref{xvarhp} must be expressible as a total derivative if
one uses the equations of motion. It is not immediately obvious how to do this, but the derivation in section \ref{result} suggests 
a strategy. One can consider the ansatz and scale $x^1\to x^1/k$ to place all of the $k$ dependence into $h$. After some work one finds that
\begin{align}\label{exintegral}
k\frac{\delta w}{\delta k} &=\int_{r_+}^\infty dr\left[\xi_1-2r^2fgh'\right]'
-\lim_{r\to\infty}\left(\ln r\frac{fg^{1/2}e^{2\alpha}b^2}{h}k^2\right)_{OS}\,,
\end{align}
where $\xi_1$ was given in \eqref{xiexps}. 
The first term in the integrand arises from the fact that we are considering the variation of a density. To obtain the second term we observe that the scaling $x^1\to x^1/k$ leads to
$h\to h/k\equiv\tilde h$. Thus, as a perturbation of the rescaled metric we have
$\delta \tilde g_{x^1 x^1}=-2\tilde h^2\delta k/k$. Now varying the
bulk action \eqref{eq:lagemcs} we will obtain boundary terms only from the variation of
the Einstein-Hilbert term\footnote{If we had an ansatz in which there was also $k$ dependence appearing in the gauge field $A$ there would
be additional contributions from the gauge kinetic term and also the Chern-Simons term.} via
\begin{align}
\delta(\sqrt{-g}R)&=\sqrt{-g}\nabla^\mu\left(\nabla^\nu\delta g_{\mu\nu}-g^{\rho\sigma}\nabla_\mu\delta g_{\rho\sigma}\right)\,,\nn
&=\partial_r(2r^2fgh')\delta k/k\,,
\end{align}
where we dropped the tildes in the first line for clarity. The opposite sign appearing in \eqref{exintegral} arises for the Euclidean action. 
After substituting the expansions \eqref{mubexpmet}, \eqref{nhexp} we find that
the integral on the right-hand side of expression \eqref{exintegral} only gets a contribution at $r\to\infty$, leading to
the result \eqref{ldg} which we obtained from more general arguments above. Note that in the case that
$\mu_b=0$, the divergences arising in the two terms in the integral in \eqref{exintegral}
cancel and the final boundary term does not
contribute.

Finally, we can easily check that the constraints \eqref{consone}, \eqref{constwo}, 
which in the present context read
\begin{align}\label{constsexin}
\int_{0}^{L_{1}}dx^1\left(-T^{x^it}+J^{x^i} \mu\right)=0\,,&\nn
\int_{0}^{L_{1}}dx^1\left(T^{x^ix^j}+J^{x^i} a_{x^j}\right)=0&,\qquad i\ne j
\end{align}
are indeed satisfied.

\section{Other examples}\label{othex}
Let us briefly discuss three other examples that have been considered in the literature.
The examples consist of CFTs in flat spacetime at finite temperature $T$ and constant 
chemical potential $\mu$, with no other deformations, which spontaneously undergo a phase 
transition at some critical temperature. In the first two examples the new phase spontaneously breaks translation invariance in one direction, 
while in the third it doesn't. We should note that the examples have additional matter content than we have ben considering
but the main results we mention below do not depend on the details.

The deformations of the CFT of interest imply that $\gamma_{\mu\nu}=\eta_{\mu\nu}$, $a_t\equiv \mu$ and $a_{x^i}=0$.
Any spatial modulation is again taken to be just in the $x^1$ direction, with period $L_1=2\pi/k$.
If we invoke conservation of the stress tensor we have $\partial_\mu T^{\mu\nu}=\partial_{x^1}T^{x^1\nu}=0$
and hence
\begin{align}\label{Teecons}
T^{x^1t}, \quad T^{x^1x^1},\quad\dots ,\quad T^{x^1x^{d-1}}\qquad\text{are constants}\,.
\end{align}
Similarly, conservation of the current, $\partial_\mu J^\mu=0$ implies that 
\begin{align}
J^{x^1}\qquad\text{is constant}\,.
\end{align}
The results \eqref{eq:final_var2}, \eqref{wst} then imply (setting $\delta\gamma_{\mu\nu}=\delta a_{x^i}=0$ for simplicity)
\begin{align}\label{eq:final_var_ill}
\delta w&=-\bar J^{t}\delta \mu-s{\delta T}+\frac{\delta k}{k}\left(w+T^{x^1x^1}\right)\,,\notag\\
w&=-\bar T^{x^2x^2}=\dots=-\bar T^{x^{d-1}x^{d-1}}\,,\nn
w&=-Ts-\bar J^{t}\mu+\bar T^{tt}\,,
\end{align}
where the bars refer to quantities averaged in the $x^1$ direction;
$\bar J^t=(k/2\pi)\int_0^{2\pi/k} dx^1 J^t$ and $\bar T^{x^ix^i}=(k/2\pi)\int_0^{2\pi/k} dx^1 T^{x^ix^i}$. 
In the case that $x^1$ is non-compact, or when there is no dependence on $x^1$,
we should impose $\delta w/\delta k=0$ at fixed $T,\mu$
to obtain the thermodynamically preferred black holes, which is equivalent to $w=-T^{x^1x^1}$. These black holes\footnote{It is worth noting for these black holes that we have, for example, $\delta T^{x^1x^1}=-\delta w-\bar J^{t}\delta \mu+s{\delta T}$, similar to 
\cite{Traschen:2001pb}.}
can be equivalently characterised by demanding that 
the averaged pressures in all directions are equal: $T^{x^1x^1}=\bar T^{x^2x^2}=\dots=\bar T^{x^{d-1}x^{d-1}}\equiv\bar p$. We also have $w=-\bar p$ and 
\begin{align}
\bar T^{tt}+\bar p=Ts+\bar J^{t}\mu\,.
\end{align}
For the backgrounds we are considering, the stress tensor is traceless, 
$T^{\mu}{}_\mu=0$, which also implies that $\bar T^{tt}=(d-1)\bar p$. 
The constraint \eqref{consone} now implies
\begin{align}\label{constsex}
\int_{0}^{L_{1}}dx^1\left(-T^{x^it}+J^{x^i} \mu\right)=0\,,
\end{align}
while the constraint \eqref{constwo}, when combined with \eqref{Teecons}, implies
\begin{align}\label{constsex2}
\int_{0}^{L_{1}}dx^1\left(T^{x^ix^j}\right)=0&,\qquad i\ne j\ne 1,\nn 
T^{x^1x^i}=0&,\qquad i\ne1\,.
\end{align}

We now turn to the three examples. In \cite{Donos:2012gg} $D=5$ black hole solutions were constructed that are dual to 
$p$-wave helical superconductors. The axis of the helix is along the $x^1$ direction, with the solutions
being spatially modulated in the $x^1$ direction and translationally invariant in the $x^2$ and $x^3$ directions.
The boundary stress tensor and current were obtained for these black hole solutions and
after substituting into \eqref{eq:final_var_ill} one finds results for $w$ and $\delta w$ that precisely agree with
those given in \cite{Donos:2012gg}. In particular, in the non-compact case, the condition 
$T^{x^1x^1}=\bar T^{x^2x^2}=\bar T^{x^3x^3}$ for the thermodynamically preferred black 
holes that we have just obtained,
translates in the language of \cite{Donos:2012gg} to the condition that $c_h=0$;  this is precisely the
condition that was found from the numerical calculations. Furthermore, the solutions have 
$T^{tx^i}=J^{x^i}=T^{x^1x^2}=T^{x^1x^3}=0$
and hence the only non-trivial content of \eqref{constsex}, \eqref{constsex2} is that the integral of $T^{x^2x^3}$ along a period of 
$x^1$ should vanish, which indeed it does for the solutions of \cite{Donos:2012gg}.

Similarly, we can also compare \eqref{eq:final_var_ill} with the results obtained for the $D=4$
striped black hole solutions found in \cite{Donos:2011bh,Rozali:2012es,Donos:2013wia,Withers:2013loa,Withers:2013kva,Rozali:2013ama}. 
In particular, the boundary stress tensor and current were given in 
\cite{Donos:2013wia,Withers:2013kva}. 
The black hole solutions are spatially modulated in the $x^1$ direction and translationally invariant
in the $x^2$ direction.
For the non-compact case, 
the relation $T^{x^1x^1}=\bar T^{x^2x^2}$ for the thermodynamically preferred black 
holes precisely agrees with the numerical results of \cite{Withers:2013kva}.

The last example concerns the D=5 $p$-wave superconducting black holes that have been constructed in \cite{Basu:2009vv,Ammon:2009xh}, building on \cite{Gubser:2008wv}. These solutions are {\it not} spatially modulated but they have an anisotropic structure associated with the
$p$-wave order.
All components of the stress tensor are constant for these solutions and hence our general results
imply that the stress tensor must be spatially isotropic, $T^{x^1x^1}=T^{x^2x^2}= T^{x^{3}x^{3}}$, in apparent contrast
to the results of \cite{Ammon:2009xh}. However, closer inspection shows that consistency can be achieved if
the constant $f_2^b$ appearing in equation (21) of \cite{Ammon:2009xh}
is actually equal to zero. We have confirmed with some of the authors of \cite{Ammon:2009xh} that
their numerical solutions do in fact have $f_2^b=0$, up to numerical error \cite{privcom}.

\section{Final Comments}\label{fincom}
In essence, our main results in section two arose from the holographic result \eqref{firstvar} 
combined with scaling arguments and reparametrisations of the boundary torus.
While we focussed on CFTs with a bulk theory containing a metric and a gauge-field, the generalisation
to other matter fields is straightforward. Another generalisation is to incorporate applied magnetic fields. 
Indeed there has been various investigations into spatially modulated CFTs in the presence of magnetic fields including
\cite{Maeda:2009vf,Ammon:2011je,Almuhairi:2011ws,Donos:2011qt,Donos:2011pn,Donos:2012yu,Almheiri:2011cb,Bu:2012mq,Bao:2013fda}.
For simplicity, we assume that there is magnetic field in the $x^1,x^2$ plane so that
\eqref{btwo} is replaced with
\begin{align}\label{btwof}
A_B&=a_{\mu}\,dx^{\mu}+\tfrac{1}{2}B(x^1dx^2-x^2dx^1)\,,
\end{align}
with $a_\mu$ periodic functions of $x^{i}$ only. The variation of the action given by \eqref{firstvar} will pick up an extra term given by
$-\Delta \tau \Pi m\delta B$, where $m$ is, by definition, the magnetisation per unit volume.
We then find that, for example, the variation of the free energy density given in 
\eqref{eq:final_var2} should be replaced with
\begin{align}
\delta w&=
-\Pi^{-1}\int_{0}^{\{L_{i}\}}d^{d-1}x\,\sqrt{-\det{\gamma}}\,\left(\tfrac{1}{2}T^{\mu\nu}\,\delta \gamma_{\mu\nu}+J^{\mu}\,\delta a_{\mu} \right)-s{\delta T}-m\delta B\notag\\
&-\sum_{i,j}\frac{\delta L_{i}}{L_{i}}\left[w+(\delta^{i1}+\delta^{i2})mB
+\Pi^{-1}\,\int_{0}^{\{L_{i}\}}d^{d-1}x\sqrt{\det{\gamma}}\,\left(T^{tx^i}\gamma_{tx^i}+T^{x^ix^j}\gamma_{x^ix^j}+J^{x^i}a_{x^i} \right)\right]\,,
\end{align}
and the entropy density is unchanged from that derived from \eqref{strop}. 

Our results will also generalise to other holographic settings where the AdS asymptotic behaviour is 
replaced with asymptotic behaviour associated with theories exhibiting, for example,
Lifshitz scaling \cite{Kachru:2008yh} or hyperscaling violation \cite{Charmousis:2010zz,Ogawa:2011bz,Huijse:2011ef}. This will be relevant to the thermodynamics of the 
spatially modulated configurations that have been studied in 
\cite{Cremonini:2012ir,Iizuka:2013ag,Donos:2013gda,Bao:2013ixa}.

\section*{Acknowledgements}
We thank J. Erdmenger, D. Kastor, C. Pantelidou, K. Skenderis, A-K. Straub, M.Taylor, J. Traschen, B. Withers and H. Zeller for helpful discussions. 
The work is supported in part by STFC grant ST/J0003533/1.


\providecommand{\href}[2]{#2}\begingroup\raggedright\endgroup

\end{document}